\documentclass[10pt,a4paper]{article}
\usepackage{amsmath,amssymb,epsfig,float,subfig,cite}

\newcommand{\EM}{electromagnetic}
\newcommand{\BAI}{Born and Infeld}
\newcommand{\BI}{Born-Infeld}
\newcommand{\ph}{ \varphi}
\newcommand{\PD}{ {\partial} }
\newcommand{\hatl}{{\hat\lambda  }}
\newcommand{\DD}{{ \sqrt{\Delta( \ph,\Lambda,  {\cal A}  )} }}
\newcommand{\DDD}{{ \sqrt{\Delta( \ph(\rho_0),\Lambda(\rho_0),  {\cal A}(\rho_0)  )} }}

\title{Born-Infeld Axion-Dilaton Electrodynamics and Electromagnetic Confinement}

\author{D.A. Burton\thanks{Physics Department, Lancaster University, Lancaster LA1 4YB, UK}
\thanks{The Cockcroft Institute of Accelerator Science and Technology, Daresbury WA4 4AD, UK}
\and
T. Dereli\thanks{Department of Physics, Ko\c{c} University, 34450 Istanbul, Turkey}
\and
R.W. Tucker\footnotemark[2]
\footnotemark[1]}

\begin{document}

\maketitle

\section{Introduction}
The existence of new forms of matter that interact only with gravitation has been recently advocated in order to account for a number of puzzles in modern cosmology. However the experimental detection of such states remains elusive. Unified models of the basic interactions also predict a large class of undetected states that may induce experimental signatures predicted by low energy effective string models. Phenomenological models of the strong interactions (QCD) also demand ``axionic'' states to ameliorate anomalies in the presence of the observed lepton families and account for the observed imbalance of matter over anti-matter~\cite{peccei:1977, weinberg:1978, wilczek:1978}.  Furthermore the simplest generalization of  Einsteinean gravitation involves a gravitational scalar field that modifies certain predictions of Einstein's theory~\cite{brans:1961, td_rwt:2002}. Perhaps the coupling of hypothetical axions and dilaton scalar fields to the \EM{ }field offers the most promising mechanism leading to their experimental detection~\cite{ahlers:2008}. It is therefore worth analyzing new effective field theories involving such interactions~\cite{noble:2008}. Although a number of traditional searches for axion particles are based on natural modifications to the linear Maxwell theory in vacuo, this may be a weak-field approximation to a more general non-linear vacuum electrodynamics. Indeed, in the absence of axions and dilatons, such a theory was first formulated by \BAI{ }~\cite{born:1934} in 1934. This theory has acquired a modern impetus from the observation that it emerges naturally in certain string-inspired quantum field theories~\cite{fradkin:1985} and it is perhaps unique among a large class of non-linear electrodynamic models in its causal properties in background spacetimes~\cite{boillat:1970, plebanski:1968}. String theories also naturally include candidates for axion and dilaton states that at the Planck scale have prescribed couplings among themselves and the Maxwell field. In low-energy effective string models these couplings give rise to particular symmetries in the weak-field limit. Such models have been extensively studied by Gibbons et al~\cite{gibbons:2001, gibbons:1995, gibbons:1996} with particular reference to the preservation of linear realizations of ${\rm SL}(2,\mathbb{R})$ symmetry~\cite{gibbons:1995} and non-linear realizations of \EM{ }duality in the context of \BI{ }electrodynamics with a dilaton~\cite{gibbons:1996}. In this letter we report on a new model that naturally incorporates both axion and dilaton fields in the context of \BI{ }vacuum non-linear electrodynamics.
\section{Axion-dilaton \BI{ }Electrodynamics}
If $\{R^a{ }_b\}$ denotes the curvature 2-forms of the Levi-Civita connection, $g =\eta_{ab}\, e^a \otimes e^b$ the spacetime metric  with $\eta_{ab}= diag\{ -1,1,1,1 \}$ and $\{e^a\}$ a local $g$-orthonormal co-frame, $F$ the Maxwell 2-form, $\varphi$ the dilaton scalar and ${\cal A}$ the axion scalar, the model that arises from string theory in a weak-field limit~\cite{gibbons:1995} follows by a variation of the action  $S[g,A,\ph,{\cal A}]= \int_M \Lambda_0$  where the 4-form $\Lambda_0$ on spacetime $M$  is
\begin{align}
\notag
\Lambda_0 =& p_1\, R_{ab}\wedge\ast\, (e^a\wedge e^b) + p_2\,d\varphi\wedge\ast\, d\varphi
+ p_3\,\exp(-2\varphi)\,d{\cal A}\wedge\ast\, d{\cal A}\\
&+ p_4\,{\cal A}\,F\wedge F + p_5\,\exp(\varphi)\, F\wedge\ast F\label{act1}
\end{align}
with $F=d\,A$ and $\ast$ is the Hodge map associated with $g$. In (\ref{act1}) the constants are
\begin{align}
&p_1 = \frac{c^3}{8\pi G_N}=\frac{\hbar}{L^2},\\
&p_2 = p_3 = -\frac{2\hbar}{L^2},\\
&p_4 = p_5 = \frac{\varepsilon_0}{2c},
\end{align}
in terms of the Planck length
\begin{equation}
\label{Planck_length}
L = \sqrt{\frac{8\pi \hbar G_N}{c^3}},
\end{equation}
the Newtonian gravitational constant $G_N$ and the permittivity of free space $\epsilon_0$~
\footnote{All tensor fields in this article have dimensions
  constructed from the SI dimensions $[M], [L], [T], [Q]$ where $[Q]$
  has the unit of the Coulomb in the MKS system. We adopt $[g]=[L^2],[\varphi]=[{\cal A}  ]=1,
  [G]=[Q],\,[F]=[Q]/[\epsilon_0]$ where the permittivity of free space
  $\epsilon_0$ has the dimensions $ [ Q^2\,T^2 M^{-1}\,L^{-3}] $ and
  $c=\frac{1}{\sqrt{\epsilon_0\mu_0}}$ denotes the speed of light in vacuo.
  Note that, with $[g]=[L^2]$, for $r-$forms $\alpha$ in $4$ dimensions one has $[\ast \alpha]=[\alpha] [L^{4-2r}]$.}.  The term involving $p_4$ in (\ref{act1}) denotes the traditional coupling of the axion field to the \EM{ }field while the term involving $p_5$ is a natural dilaton coupling.
One of the original aims given by \BAI{ }in generalizing the vacuum Maxwell theory was to construct a theory possessing bounded spherically symmetric static electric fields. Their theory invoked a new fundamental constant $b_0$ with the physical dimensions of an electric field strength.  They demonstrated that their field equations  admitted such solutions. Furthermore, by assuming that the finite mass of such an \EM{ }field configuration could be identified with the electron Born and Infeld were able to estimate the magnitude of $b_0$.  While such an argument is suspect in the context of subsequent developments, the idea of ameliorating the Coulomb singularity in the electric field using a non-linear \EM{ }self-coupling remains attractive.

The generalization considered here also reduces to the model defined by (\ref{act1}) in a weak-field limit. Furthermore in the absence of axion and dilaton contributions it reduces in the limit $b_0 \mapsto \infty$ to  Maxwell's  linear vacuum theory while for finite $b_0$ it reduces to the original \BI{ }model. It involves the Newtonian gravitational coupling constant $G_N$ in addition to $b_0$ and a parameter $\tau=\pm 1$ and is obtained by varying the action:
\begin{equation}
S_\tau[g, A,\varphi,{\cal A}] =\int_M {\Lambda}_\tau
\end{equation}
where
\begin{align}
\notag
\Lambda_\tau =&  p_1\,R_{ab}\wedge\ast(e^a\wedge e^b)
+ p_2\,[d\varphi\wedge\ast d\varphi + \exp(-2\varphi)\, d{\cal A}\wedge d{\cal A}]\\
&+ f_\tau(X,Y,\varphi,{\cal A})\ast 1,\label{act2_Einstein}
\end{align}
with
\begin{equation}
\label{f-tau_definition_Einstein}
f_\tau(X,Y,\varphi,{\cal A}) = \tau \frac{\varepsilon_0 b_0^2}{c}\,\bigg\{1-\sqrt{1-\tau\frac{e^\varphi\,X}{b_0^2}
-\tau\frac{{\cal A} Y}{b_0^2} - \frac{[e^\varphi\,Y - {\cal A} X]^2}{4 b_0^4}}
\,\bigg\},
\end{equation}
$ X=\ast(F \wedge \ast F )  $ and $Y= \ast( F \wedge F  )   $.

The structure of the argument of the square root in (\ref{f-tau_definition_Einstein}) follows from the (non-trivial) identity
\begin{align}
\notag
-det&\bigg( \eta_{ab} + \frac{\sqrt{\tau}}{b_0}\alpha F_{ab}   +    \frac{\sqrt{\tau}}{b_0}\beta \tilde F_{ab}  \bigg)\\
&= 1-\tau\exp(\ph)\, \frac{ X} { b_0^2  }     - \tau\, {\cal A} \frac{Y  }  { b_0^2  }  - \frac{[\exp(\ph) Y - {\cal A} X ]^2 }{ 4 b_0^4  }
\end{align}
\noindent where $\exp(\ph)= \alpha^2-\beta^2$, ${\cal A}=-2\alpha\beta$,   $F= \frac{1}{2} F_{ab} e^a \wedge e^b$   and $\ast F= \frac{1}{2} \tilde F_{ab} e^a \wedge e^b$.

The $4$-form $\Lambda_\tau$ in (\ref{act2_Einstein}) is expressed in the so-called Einstein ``frame''. However, it turns out that the importance of the dilaton and axion in our model is more readily exposed by moving to the Jordan ``frame'' by making the Weyl transformation $g \mapsto {\bf g} = \psi^{-2}\,g$ where $\psi=\exp(\varphi)$. After the field equations have been obtained in the following by varying $\Lambda_\tau$,  we will choose ${\bf g}$ to be a flat background metric thereby endowing the metric $g$ in the Einstein frame with non-zero curvature where the dilaton is non-constant.

Introducing ${\bf e}^a = \psi^{-1}\,e^a$ where $\{{\bf e}^a\}$ is a ${\bf g}$-ortho-normal co-frame, it follows that
\begin{align}
\notag
\Lambda_\tau =& p_1\,\psi^2\,{\bf R}_{ab}\wedge\star ({\bf e}^a \wedge {\bf e}^b) + (6 p_1 + p_2)\,d\psi\wedge\star d\psi
+ p_3\,d{\cal A}\wedge\star d{\cal A}\\
&+{\bf f}_\tau({\bf X},{\bf Y},\psi,{\cal A})\star 1
\label{act1_2}
\end{align}
with $\star$ the Hodge map associated with ${\bf g}$, $\{{\bf R}^a{ }_b\}$ the curvature $2$-forms of the Levi-Civita connection of ${\bf g}$, ${\bf X} = \star(F\wedge\star F)$, ${\bf Y} = \star(F\wedge F)$ and
\begin{equation}
\label{f-tau_definition_Jordan}
{\bf f}_\tau({\bf X},{\bf Y},\psi,{\cal A}) = \tau \frac{\varepsilon_0 b_0^2}{c}\,\psi^4\bigg\{1-\sqrt{1-\tau\frac{{\bf X}}{b_0^2\,\psi^3}
-\tau\frac{{\cal A}{\bf Y}}{b_0^2\,\psi^4} - \frac{[\psi\,{\bf Y} - {\cal A} {\bf X}]^2}{4 b_0^4\,\psi^8}}
\,\bigg\}.
\end{equation}
In the following attention is restricted to the case where $\tau=1$.

The non-linear vacuum Maxwell equations follow as $d\, F=0$ (since $F=d\,A$) and (by varying $A$)
  \begin{equation}
\label{maxwell_in_Jordan}
  d\,\star G=0
  \end{equation}
  where
  \begin{equation}
\label{G_in_Jordan}
\star G = 2 c\,{\bf f}_{\bf X} \star F + 2 c\,{\bf f}_{\bf Y} F
\end{equation}
and ${\bf f}_{\bf X}=\partial_{\bf X}\,{\bf f}$ etc.
From ${\cal A}$ variations one has
\begin{equation}
\label{axion_eqn_in_Jordan}
-2 p_3\, d\star d{\cal A} + {\bf f}_{\cal A}\star 1 = 0
\end{equation}
and from $\psi$ variations
\begin{equation}
\label{dilaton_eqn_in_Jordan}
-2 (6 p_1 + p_2)\,d\star d\psi + 2 p_1\,\psi\,{\bf R}_{ab}\wedge\star({\bf e}^a\wedge {\bf e}^b) + {\bf f}_\psi \star1 = 0.
\end{equation}
Using ${\bf g}$-ortho-normal co-frame variations one obtains the gravitational field equations in  the Jordan frame
\begin{equation}
\label{Einstein_eqn_in_Jordan}
p_1\,\psi^2\,{\bf R}^{bc}\wedge \star({\bf e}_a\wedge {\bf e}_b \wedge {\bf e}_c) = \tau_a [F,\psi,{\cal A},g]
\end{equation}
where
\begin{align}
\notag
\tau_a [F,\psi,{\cal A},g] =\,&  (6 p_1 + p_2) ( i_{{\bf X}_a} d\psi\wedge \star d\psi + d\psi \wedge i_{{\bf X}_a}\star d\psi)\\
\notag
+ &p_3 (i_{{\bf X}_a} d{\cal A}\wedge \star d{\cal A} + d{\cal A}\wedge i_{{\bf X}_a}\star d{\cal A})\\
\label{stress_forms_in_Jordan}
- &({\bf f} - {\bf X}\,{\bf f}_{\bf X} - {\bf Y}\,{\bf f}_{\bf Y})\star {\bf e}_a - {\bf f}_{\bf X} (i_{{\bf X}_a} F\wedge\star F - F\wedge i_{{\bf X}_a}\star F).
\end{align}
Solving the full field systems in the Einstein and Jordan frames should lead to the same conclusions (up to identification of the spacetime metric) but finding exact solutions to such systems is non-trivial.
A natural approximation is to neglect the couplings of dynamic gravitation to the other fields by neglecting the gravitational field equations. However the neglect of these equations in the Einstein frame will, in general, lead to solutions for the axion, dilaton and electromagnetic fields in a $g$-flat background with different behaviours from those calculated in a ${\bf g}$-flat background in the Jordan frame.

To probe the consequences of adopting a ${\bf g}$-flat background metric in the presence of the explicit coupling of the dilaton $\psi$ to the curvature ${\bf R}^a{ }_b$ in (\ref{dilaton_eqn_in_Jordan}) we develop two  approximation schemes in the Jordan frame.   The first approach neglects  (\ref{Einstein_eqn_in_Jordan}) from the outset with ${\bf g}$  set  to a background ${\bf g}$-flat metric. The second approach  employs (\ref{Einstein_eqn_in_Jordan}) to express the Ricci scalar in terms of the dilaton field.   The result is used to eliminate ${\bf R}_{ab}\wedge\star({\bf e}^a\wedge {\bf e}^b)$ from (\ref{dilaton_eqn_in_Jordan}), yielding
\begin{align}
\label{trace_eqn}
p_2\, d\star d \psi^2 + 2 p_3\, d{\cal A}\wedge\star d{\cal A} - \psi\,{\bf f}_\psi \star 1
+ 4 ({\bf f} - {\bf X}\,{\bf f}_{\bf X} - {\bf Y}\,{\bf f}_{\bf Y}) \star 1 = 0.
\end{align}
We then  set ${\bf g}$ to the background ${\bf g}$-flat metric (thereby ignoring the gravitational field equation (\ref{Einstein_eqn_in_Jordan}) for the remainder of the analysis). In the following, we  refer to the field system (\ref{maxwell_in_Jordan}), (\ref{G_in_Jordan}), (\ref{axion_eqn_in_Jordan}), (\ref{dilaton_eqn_in_Jordan}) as {\it System 1} and the field system (\ref{maxwell_in_Jordan}), (\ref{G_in_Jordan}), (\ref{axion_eqn_in_Jordan}), (\ref{trace_eqn}) as {\it System 2}.

In the static spherically symmetric sector an ortho-normal co-frame field for a background metric ${\bf g}$ is ${\bf e}^0=c\,d\,t,  {\bf e}^1= d\,r,  {\bf e}^2=r\,d\,\theta,  {\bf e}^3=r\,\sin\theta\,d\,\phi$ in spherical polar coordinates. In terms of the dimensionless radial coordinate  $\rho=r/L$ we write:
\begin{equation}
F=b_0 \Lambda(\rho)\, d\,r \wedge c d\, t
\end{equation}
with $\psi=\psi(\rho), {\cal A}={\cal A}(\rho)$.  Equations (\ref{axion_eqn_in_Jordan}), (\ref{maxwell_in_Jordan}) reduce to
\begin{align}
\label{axion_DE}
&\PD_\rho^2{\cal A} + \frac{2}{\rho} \, \PD_\rho{\cal A}=  - \frac{\hatl} {8}  \frac{{\cal A} \Lambda^4} { \DD },\\
\label{constraint}
&\frac{ \rho^2 \,( 2\psi^5\Lambda + \Lambda^3 {\cal A}^2   )  }  {4 \DD   }=\Gamma_0
\end{align}
for some integration constant $\Gamma_0$.  The dimensionless constant $\hatl$ is defined by
\begin{equation}
\label{hat_lambda}
\hatl= \left(   \frac{8 \pi \hbar G_N }  { c^3}  \right)^2 \frac{ \epsilon_0 b_0^2 }  {2 \hbar c }
\end{equation} and
\begin{equation}
\Delta( \psi,\Lambda,  {\cal A}  )=\psi^8 - \psi^5  \Lambda^2 - \frac{1}{4}{\cal A}^2 \Lambda^4 .
\end{equation}

The ODE for $\psi$ in System 1 follows directly from adopting a ${\bf g}$-flat  metric in (\ref{dilaton_eqn_in_Jordan}),
\begin{equation}
\label{dilaton_DE_1}
\PD_\rho^2\psi + \frac{2\,\PD_\rho\psi}{\rho} = 2\hatl\psi^3\bigg(\frac{\psi^4 - 5\psi\Lambda^2/8}{\DD} - 1\bigg),
\end{equation}
whereas (\ref{trace_eqn}) yields:
\begin{equation}
\label{dilaton_DE_2}
\PD_\rho^2\psi + \frac{2\,\PD_\rho\psi}{\rho} +  \frac{1}{\psi}[(\PD_\rho\psi)^2 + (\PD_\rho{\cal A})^2] =
- \frac{ \hatl }{4} \frac  {\psi^4 \Lambda^2}{ \DD }.
\end{equation}
 as the ODE for $\psi$ in System 2

\begin{figure}
\centering
\subfloat[Electric axion-dilaton bound state with confined fields.]{\scalebox{0.35}{\label{subfig:bounded}\includegraphics{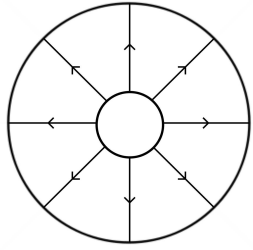}}}
\hspace{1em}
\subfloat[Electric axion-dilaton bound state with un-confined fields.]{\scalebox{0.35}{\label{subfig:unbounded}\includegraphics{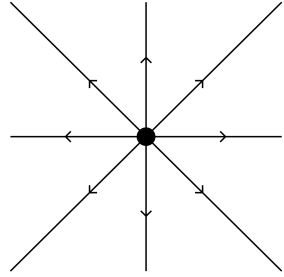}}}
\caption{\label{fig:electric_field} Examples of the electric field lines arising from numerical analyses of Systems 1 and 2. Figure \ref{subfig:bounded} illustrates a state where the electric field lines emanate and terminate on spheres of finite radius in space. The state in figure \ref{subfig:unbounded} with unconfined electric flux generalizes the static spherically symmetric solution in the original Born-Infeld electron model.}
\end{figure}
\begin{figure}
\centering
\subfloat[The solid curve is $\Lambda^2/\Lambda^2_{\rm max}$, the dashed curve is $\psi/\psi_{\rm max}$ and the dot-dashed curve is ${\cal A}/{\cal A}_{\rm max}$.]{\scalebox{0.25}{\label{subfig:fields_BI_type}\includegraphics{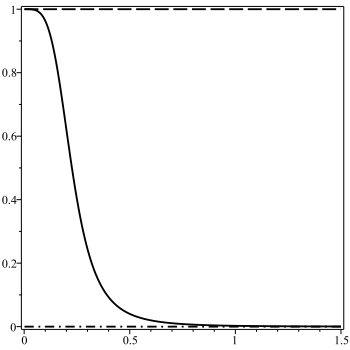}}}
\hspace{1em}
\subfloat[The solid curve is $\Gamma_0/\Gamma_{0\,{\rm max}}$ and the dashed curve is $\Delta/\Delta_{\rm max}$.]{\scalebox{0.25}{\label{subfig:gamma_delta_BI_type}\includegraphics{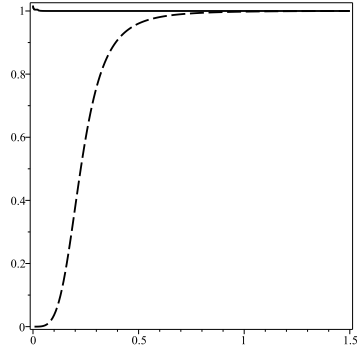}}}
\caption{\label{fig:BI_type} A solution to System 1 for which the axion and dilaton are constant and the electric field is bounded. The initial conditions are $(\psi(0.1) = 0.47, \psi^\prime(0.1) = 0, \Lambda^2(0.1) = 0.1, {\cal A}^\prime(0.1) = 0)$ and a schematic diagram of the electric field is shown in figure \ref{subfig:unbounded}. The solution to System 2 with the same initial conditions is visually indistinguishable from the above.}
\end{figure}
\begin{figure}
\centering
\subfloat[The solid curve is $\Lambda^2/\Lambda^2_{\rm max}$, the dashed curve is $\psi/\psi_{\rm max}$ and the dot-dashed curve is ${\cal A}/{\cal A}_{\rm max}$.]{\scalebox{0.25}{\label{subfig:fields_diverge_confined_zero_axion}\includegraphics{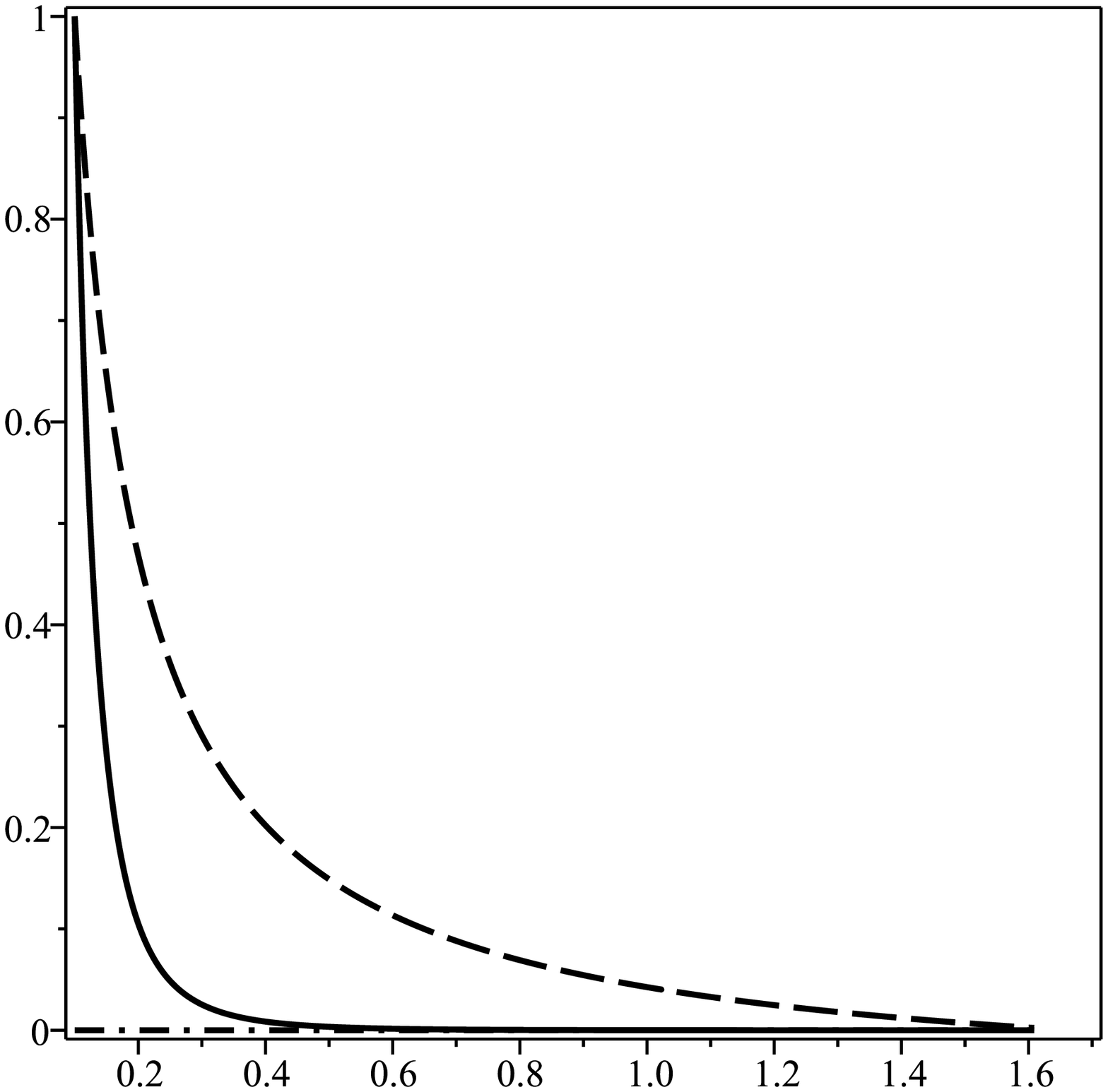}}}
\hspace{1em}
\subfloat[The solid curve is $\Gamma_0/\Gamma_{0\,{\rm max}}$ and the dashed curve is $\Delta/\Delta_{\rm max}$.]{\scalebox{0.25}{\label{subfig:gamma_delta_diverge_confined_zero_axion}\includegraphics{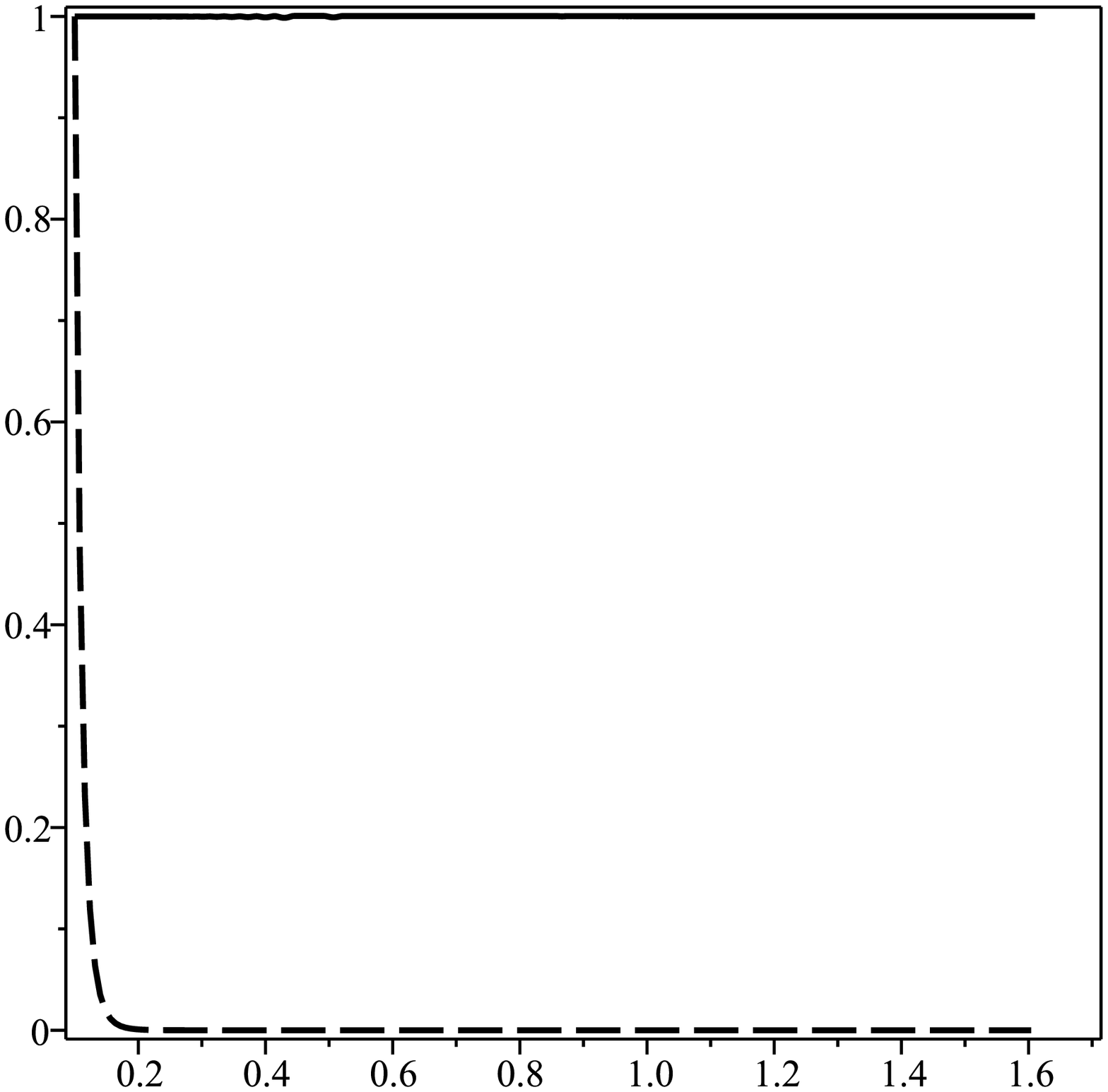}}}
\caption{\label{fig:diverge_confined_zero_axion} A solution to System 1 for which the axion, dilaton and electric fields exhibit singular behaviour for small $\rho$ and the solution terminates at $\rho = 1.61$ where $\Delta=0$. The initial conditions are $(\psi(0.1) = 0.47, \psi^\prime(0.1) = -5, \Lambda^2(0.1) = 0.1, {\cal A}^\prime(0.1) = 0)$.
}
\end{figure}
\begin{figure}
\centering
\subfloat[The solid curve is $\Lambda^2/\Lambda^2_{\rm max}$, the dashed curve is $\psi/\psi_{\rm max}$ and the dot-dashed curve is ${\cal A}/{\cal A}_{\rm max}$.]{\scalebox{0.25}{\label{subfig:stress_eliminated_fields_diverge_confined_zero_axion}\includegraphics{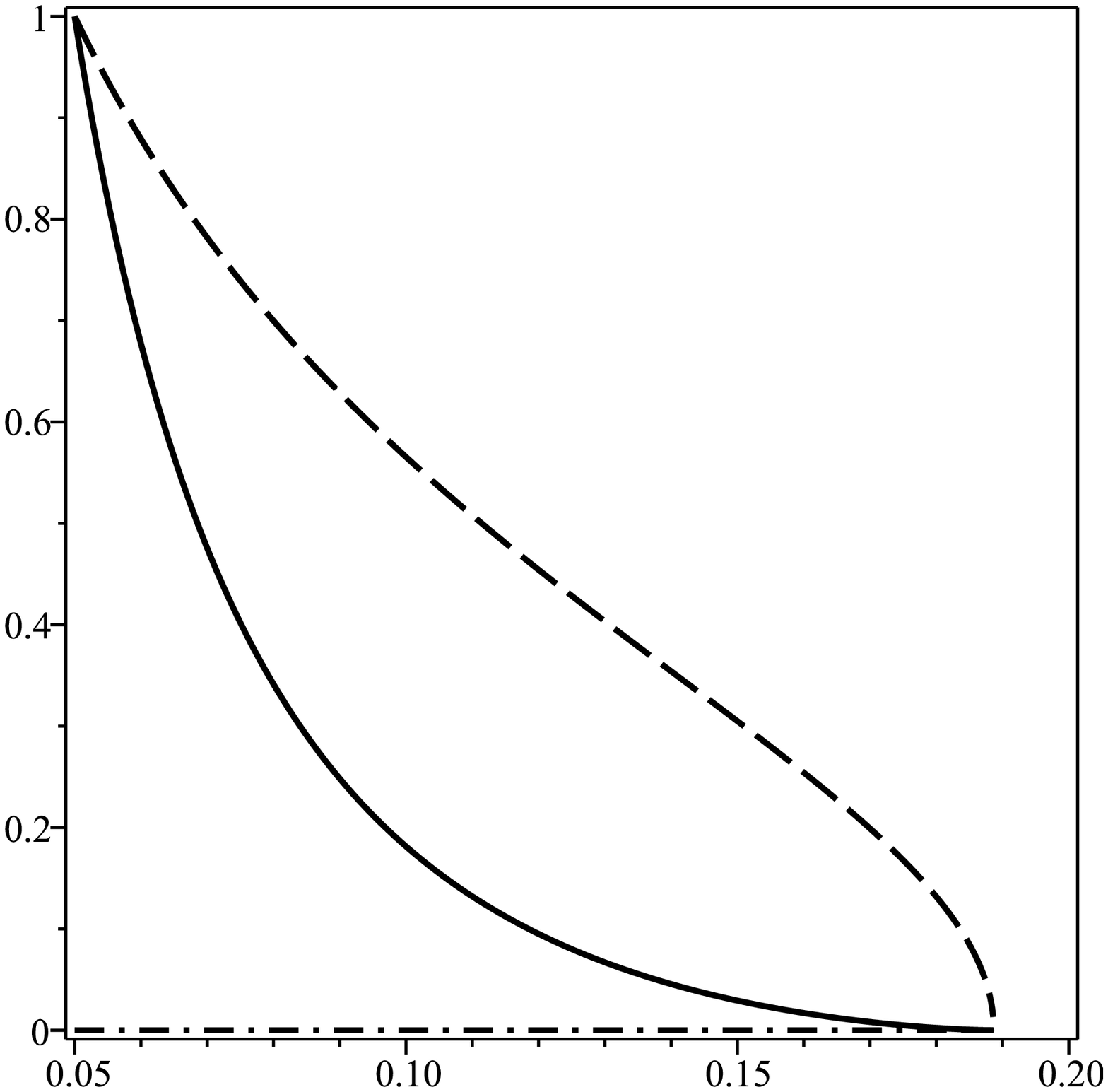}}}
\hspace{1em}
\subfloat[The solid curve is $\Gamma_0/\Gamma_{0\,{\rm max}}$ and the dashed curve is $\Delta/\Delta_{\rm max}$.]{\scalebox{0.25}{\label{subfig:stress_eliminated_gamma_delta_diverge_confined_zero_axion}\includegraphics{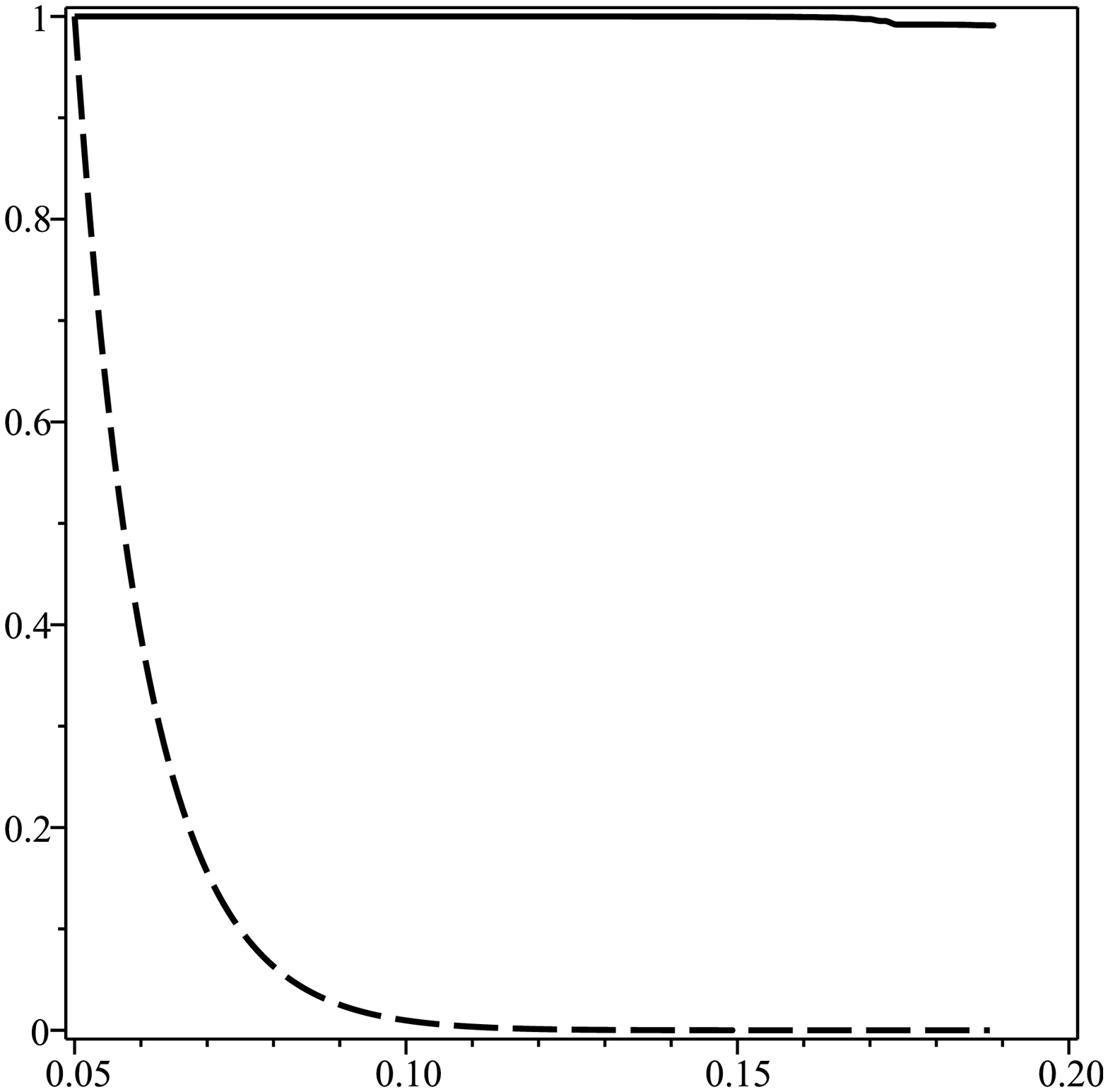}}}
\caption{\label{fig:stress_eliminated_diverge_confined_zero_axion} A solution to System 2 for which the axion, dilaton and electric fields exhibit singular behaviour for small $\rho$ and the solution terminates at $\rho = 0.189$ where $\Delta=0$. The initial conditions are the same as in figure \ref{fig:diverge_confined_zero_axion}.
}
\end{figure}
\begin{figure}
\centering
\subfloat[The solid curve is $\Lambda^2/\Lambda^2_{\rm max}$, the dashed curve is $\psi/\psi_{\rm max}$ and the dot-dashed curve is ${\cal A}/{\cal A}_{\rm max}$.]{\scalebox{0.25}{\label{subfig:fields_diverge_confined_non-zero_axion}\includegraphics{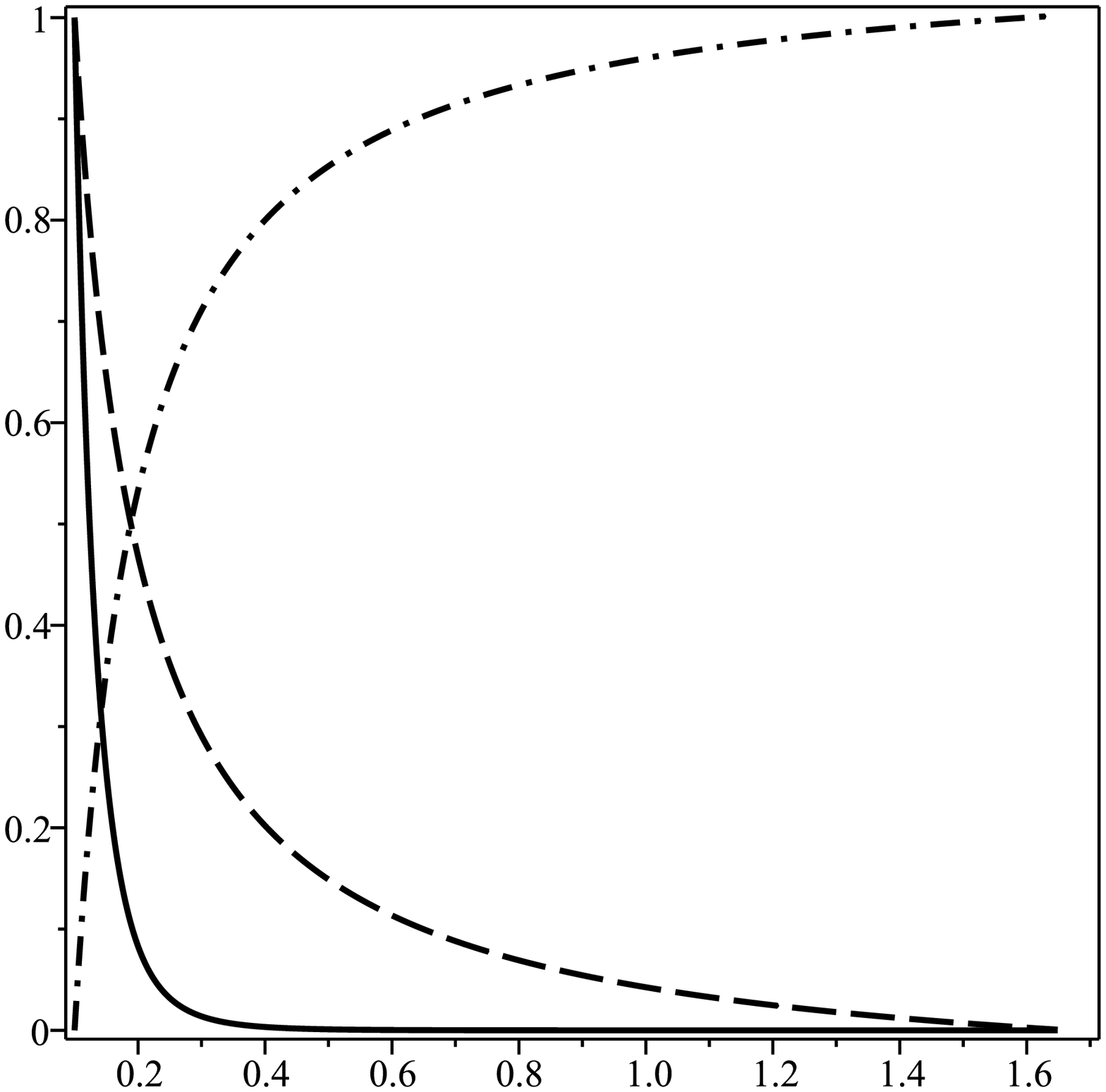}}}
\hspace{1em}
\subfloat[The solid curve is $\Gamma_0/\Gamma_{0\,{\rm max}}$ and the dashed curve is $\Delta/\Delta_{\rm max}$.]{\scalebox{0.25}{\label{subfig:gamma_delta_diverge_confined_non-zero_axion}\includegraphics{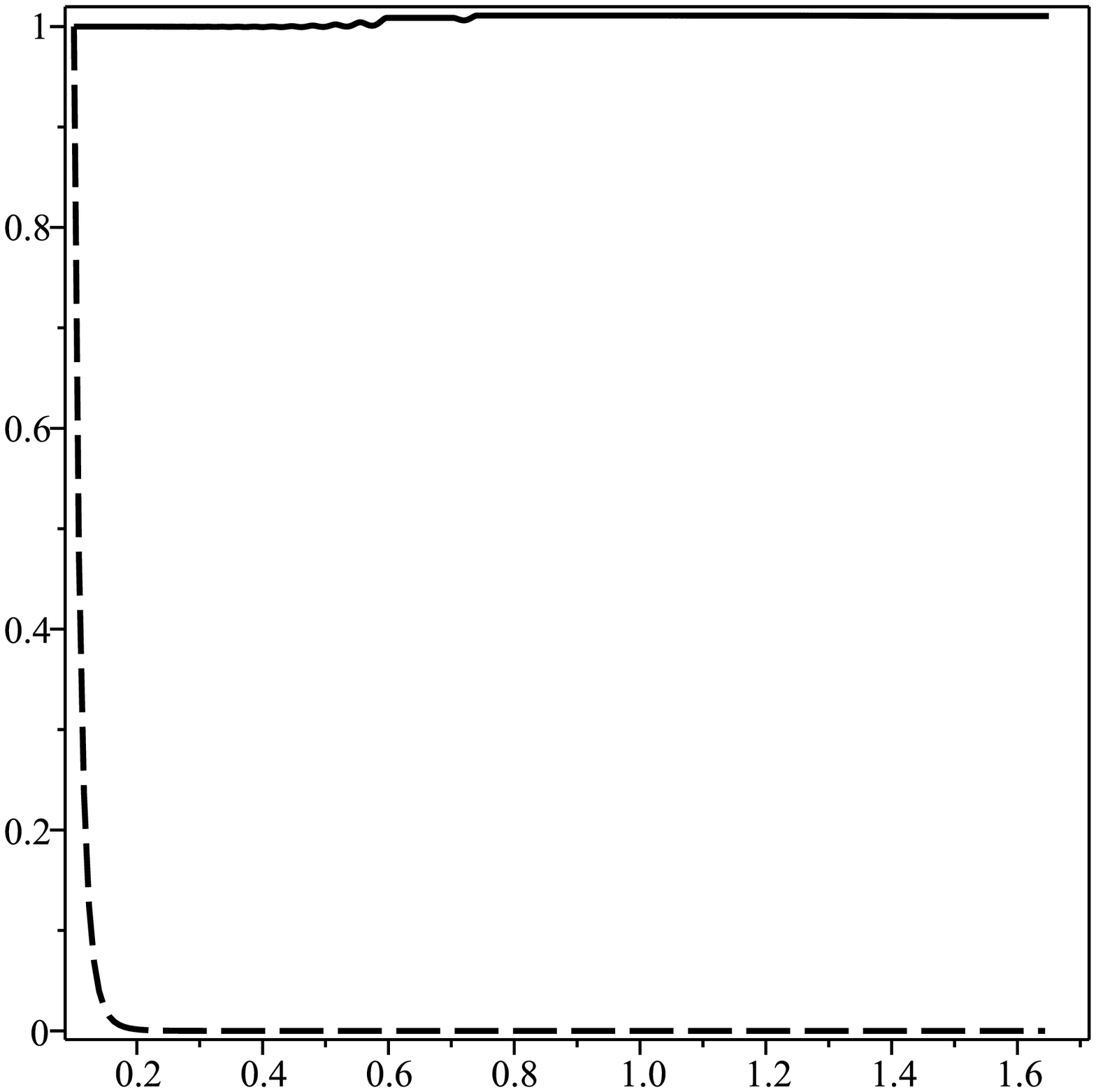}}}
\caption{\label{fig:diverge_confined_non-zero_axion} A solution to System 1 for which the axion, dilaton and electric fields exhibit singular behaviour for small $\rho$ and the solution terminates at $\rho = 1.65$ where $\Delta=0$. The initial conditions are $(\psi(0.1) = 0.47, \psi^\prime(0.1) = -5, \Lambda^2(0.1) = 0.1, {\cal A}^\prime(0.1) = 5)$. }
\end{figure}
\begin{figure}
\centering
\subfloat[The solid curve is $\Lambda^2/\Lambda^2_{\rm max}$, the dashed curve is $\psi/\psi_{\rm max}$ and the dot-dashed curve is $-{\cal A}/|{\cal A}|_{\rm max}$.]{\scalebox{0.25}{\label{subfig:fields_regular_confined_non-zero_axion}\includegraphics{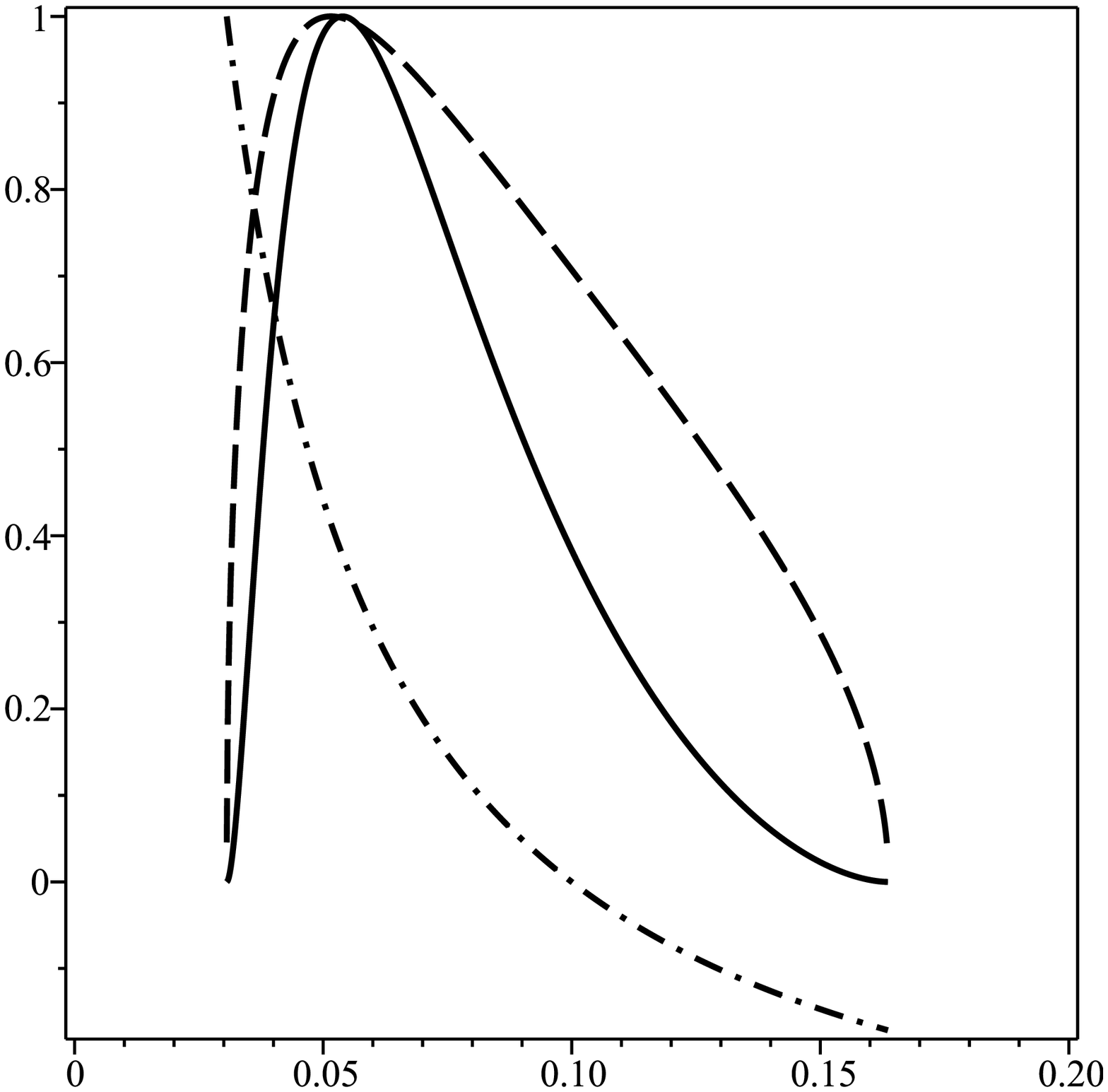}}}
\hspace{1em}
\subfloat[The solid curve is $\Gamma_0/\Gamma_{0\,{\rm max}}$ and the dashed curve is $\Delta/\Delta_{\rm max}$.]{\scalebox{0.25}{\label{subfig:gamma_delta_regular_confined_non-zero_axion}\includegraphics{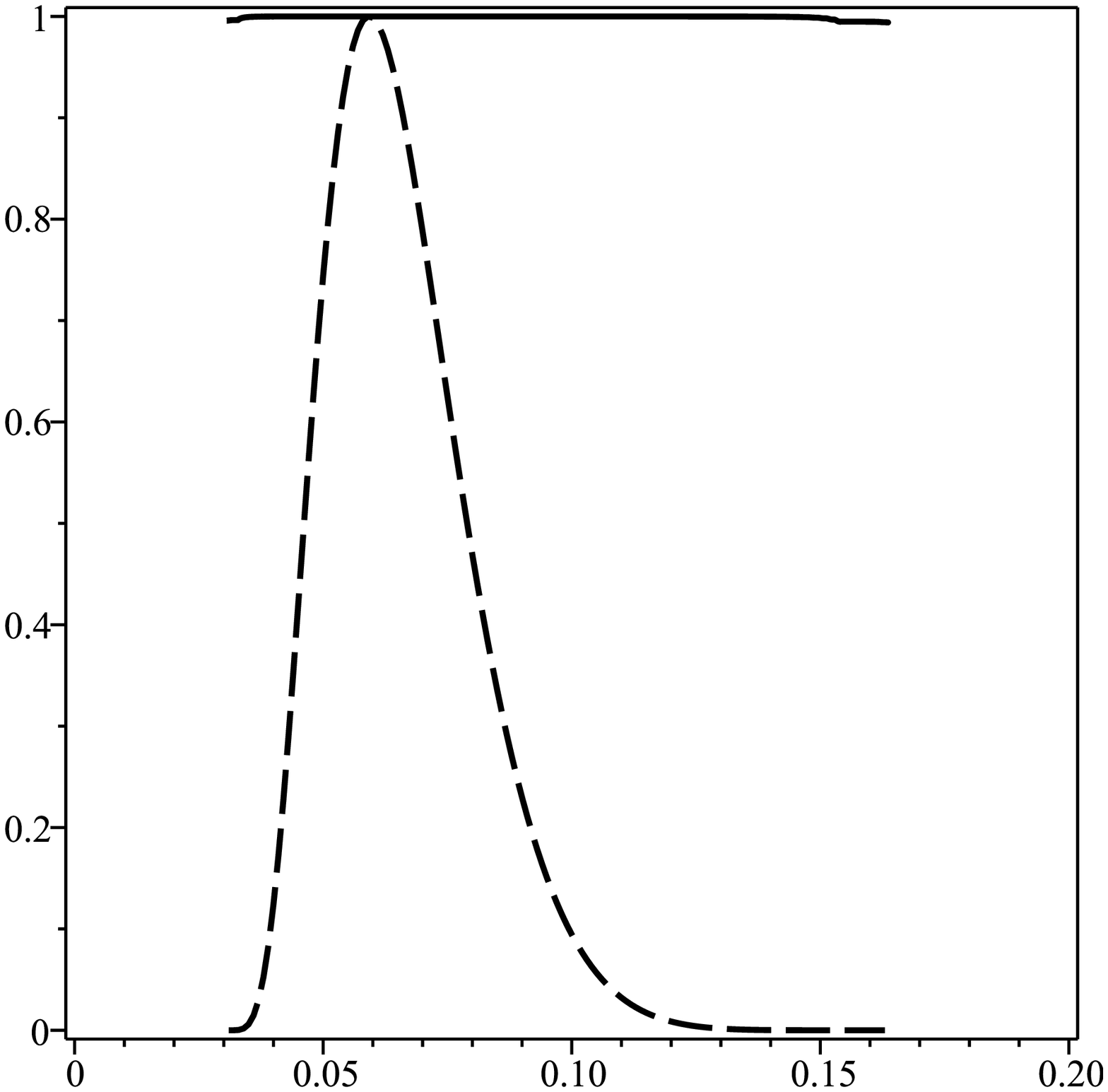}}}
\caption{\label{fig:regular_confined_non-zero_axion} A solution to System 2 for which the axion, dilaton and electric fields terminate at $\rho_a= 0.031$ and $\rho_b = 0.164$ where $\Delta=0$. The initial conditions are the same as in figure \ref{fig:diverge_confined_non-zero_axion}, and  a schematic diagram of the normalized $\Lambda^2$ determining the electric field is shown in figure \ref{subfig:bounded}. The axion field ${\cal A}$ is mostly negative over the range in $\rho$ shown above and the plot shows ${\cal A}$ normalized with its sign reversed for convenience.}
\end{figure}

If $S^2$ is any 2-sphere of radius $\rho L$ centered at $\rho=0$ the electric flux of any state crossing  the surface of this sphere is ${4\pi} q=\int_{S^2} \star G$. The value of $q$  will be interpreted as the total electric charge within this sphere.  Hence for the above spherically symmetric static field configuration determined by any constant $\Gamma_0$ such charge  is $ q= 2\epsilon_0 b_0 L^2 \Gamma_0$.

Clearly, analytic solutions to Systems 1 and 2 are unlikely; both are however amenable to numerical analysis. The simplest approach is to differentiate (\ref{constraint}) with respect to $\rho$ and treat each coupled system as an initial value problem specified by a choice of $\psi(\rho_0), {\cal A}(\rho_0), \psi^\prime(\rho_0), {\cal A}^\prime(\rho_0),  \Lambda(\rho_0)   $ with
\begin{equation}
\label{Gamma_eqn}
\Gamma_0= \frac{\rho_0^2}{4}   \left( \frac{2\psi^5(\rho_0)  \Lambda(\rho_0)  +  \Lambda^3(\rho_0)  {\cal A}^2(\rho_0)   } {\DDD } \right).
\end{equation}
The initial field conditions should be consistent with a real $\Gamma_0$.  Starting from $\rho=\rho_0$  each system can be readily integrated numerically to the regions $\rho > \rho_0$ and $\rho < \rho_0$ and the solution monitored to check that $\Gamma_0$ remains constant.

Solutions possessing the elementary charge $q=e$ were investigated and the constant $\hat{\lambda} = 10^{-80}$ was chosen, which yields a value for $b_0$ commensurate with Born and Infeld's model of the electron~\footnote{$\,\,\hat{\lambda} \sim 1$ yields a value for $\epsilon_0 b_0^2 L^3$ commensurate with the Planck energy.}. The quartic equation (\ref{Gamma_eqn}) for ${\cal A}(\rho_0)$ was used to fix the electric charge $q$ of the solution. In particular, the choice $q=e$ was implemented by algebraically solving (\ref{Gamma_eqn}) for ${\cal A}(\rho_0)$ and using the prescribed values of $\psi(\rho_0)$, $\Lambda(\rho_0)$ and $\Gamma_0 = e/ (2\epsilon_0 b_0 L^2)$. This procedure yielded ${\cal A}(\rho_0)=0$ to within numerical precision. A consistent picture that emerges from extensive numerical analysis of both systems of ODEs is the existence of {\it confined} solutions, i.e. fields that are zero for $\rho$ greater than some real positive non-zero constant.

For the same initial data  $\psi(\rho_0)$, ${\cal A}(\rho_0)$, $\ph^\prime(\rho_0)$, ${\cal A}^\prime(\rho_0)$, $\Lambda(\rho_0)$  the electric field of solutions to Systems 1 and 2 appears to be regular throughout all space (see figures~\ref{subfig:unbounded} and~\ref{fig:BI_type}; the subscript ``max'' indicates the maximum value of a field over the range of $\rho$ shown). Other solutions diverge at the origin but terminate at finite $\rho$ (see figures~\ref{fig:diverge_confined_zero_axion} to~\ref{fig:diverge_confined_non-zero_axion}). Finally, some solutions to System 2 are finite,  non-zero and continuous on a subset $\rho_a < \rho < \rho_b$  for positive non-zero $\rho_a, \rho_b$  and zero elsewhere (see figures~\ref{subfig:bounded} and \ref{fig:regular_confined_non-zero_axion}); however, we were unable to replicate this behaviour using System 1. This implies that the gravitational field equation (\ref{Einstein_eqn_in_Jordan}) plays a significant role in determining the dilaton interaction with the electromagnetic field and axion despite the neglect of  curvature in favour of a ${\bf g}$-flat metric in the Jordan frame.

The existence of states where electric,  axion and dilaton fields have finite support in space is interesting and unexpected. The role of the axion and dilaton is critical since no such configurations can occur in the spherical symmetric static sector of the original Born-Infeld theory. The field cut-offs arise when the trajectory  $\{\psi(\rho), {\cal A}(\rho), \Lambda(\rho)\}$ approaches the boundary of the domain $\DD \geq 0$ in field space.
%
\section{Conclusion}
An extension of the original \BI{ }model has been developed to include axion and dilaton fields. Motivated by low-energy effective string actions and their symmetries the model reduces in weak field or weak coupling limits to the original \BI{ }model, ${\rm SL}(2,\mathbb{R})$ covariant axion-dilaton models or linear Maxwell electrodynamics. In the absence of axion and dilaton couplings it contains only one dimensionless coupling constant $\hatl$ and is thereby analogous to the original \BI{ }model regarding its spherically symmetric static  gravity-free sector.
 Two approximation schemes  have been developed to explore the significance of dilaton couplings to curvature in the Jordan frame. Both entail working with a ${\bf g}$-flat metric although the treatment of the gravitational field equations differ in the two schemes.
 Clearly it would be valuable to find
 an exact  static spherically symmetric solution with non-zero curvature for the model in either the Einstein or Jordan frame in order to assess the validity of these schemes. If such a solution (generalizing the Reissner-Nordstr\"om solution to the Einstein-Maxwell system) exists that substantiates the approximations leading to system 2 then
 numerical evidence suggests the existence of both finite mass electrically charged and neutral states in this sector.  The latter are novel and are composed of mutually coupled electric, axionic and dilatonic fields that exist in a bounded region of space. Some states are bound by a single sphere; others are bound by two concentric spheres, much as a static electric field is confined in a spherical capacitor in Maxwell theory. The electric charge sources for such states reside in induced surface charge densities on the bounding spheres.

 The  model would then provide a mechanism for confined static abelian fields via their mutual interaction.  Since ${\rm U}(1) \subset {\rm SU}(2) \subset {\rm SU}(3)$ it would be of interest to explore whether such a mechanism arises in a non-abelian generalization.   At the abelian level it suggests the possibility of new types of electrically neutral axion-dilaton bound states with no direct interaction with external electromagnetic fields.

\section{Acknowledgments}
DAB and RWT are grateful for the hospitality provided by the Department of Physics, Ko\c{c} University
and for financial support as part of the ALPHA-X collaboration (EPSRC grant EP/E001831/1). TD is grateful for the hospitality provided by the Department of Physics, Lancaster University, and the Cockcroft Institute, UK. All authors thank TUBA (The Turkish Academy of Sciences) for
financial support.

\end{document}